%% file: derivkit.tex
\input{front_matter/preamble}

\title{\pkg{DerivKit}: stable numerical derivatives bridging Fisher forecasts and MCMC}

\date{\vspace{-7ex}}

\begin{document}
\input{front_matter/authorlist}

\maketitle

\input{front_matter/margin}
\input{front_matter/abstract}

\input{sections/intro}
\input{sections/core_functionality}
\input{sections/demo}
\input{sections/use_cases}
\input{sections/avail}
\input{sections/ack}

\bibliographystyle{mnras}
\bibliography{derivkit}

\end{document}

%% file: front_matter/preamble.tex
\documentclass[10pt,a4paper,onecolumn]{article}
\usepackage{marginnote}
\usepackage{graphicx}
\usepackage{xcolor}
\usepackage{authblk,etoolbox}
\usepackage{titlesec}
\usepackage{calc}
\usepackage{tikz}
\usepackage{hyperref}
\hypersetup{colorlinks,breaklinks=true,
            urlcolor=[rgb]{0.0, 0.5, 1.0},
            linkcolor=[rgb]{0.0, 0.5, 1.0},
            citecolor = blue}
\usepackage{caption}
\usepackage{tcolorbox}
\usepackage{amssymb,amsmath}
\usepackage{ifxetex,ifluatex}
\usepackage{seqsplit}
\usepackage{xstring}
\usepackage{natbib}
\usepackage[T1]{fontenc}
\usepackage[utf8]{inputenc}
\usepackage{cleveref}
\usepackage{float}
\usepackage{orcidlink}
\usepackage{xspace}
\usepackage{listings}
\usepackage{comment}
\let\origfigure\figure
\let\endorigfigure\endfigure
\renewenvironment{figure}[1][2] {
    \expandafter\origfigure\expandafter[H]
} {
    \endorigfigure
}

\let\textttOrig=\texttt
\def\texttt#1{\expandafter\textttOrig{\seqsplit{#1}}}
\renewcommand{\seqinsert}{\ifmmode
  \allowbreak
  \else\penalty6000\hspace{0pt plus 0.02em}\fi}

\usepackage{abstract}

\makeatletter
\let\href@Orig=\href
\def\href@Urllike#1#2{\href@Orig{#1}{\begingroup
    \def\Url@String{#2}\Url@FormatString
    \endgroup}}
\def\href@Notdoi#1#2{\def\tempa{#1}\def\tempb{#2}%
  \ifx\tempa\tempb\relax\href@Urllike{#1}{#2}\else
  \href@Orig{#1}{#2}\fi}
\def\href#1#2{%
  \IfBeginWith{#1}{https://doi.org}%
  {\href@Urllike{#1}{#2}}{\href@Notdoi{#1}{#2}}}
\makeatother

\newlength{\cslhangindent}
\newlength{\csllabelwidth}
 {
  \setlength{\parindent}{0pt}
  \ifodd #1 \everypar{\setlength{\hangindent}{\cslhangindent}}\ignorespaces\fi
  \ifnum #2 > 0
  \setlength{\parskip}{#2\baselineskip}
  \fi
 }%
 {}
\usepackage{calc}

\usepackage[top=3.5cm, bottom=3cm, right=1.5cm, left=1.0cm,
            headheight=2.2cm, reversemp, includemp, marginparwidth=4.5cm]{geometry}

\titleformat{\section}
  {\normalfont\sffamily\Large\bfseries}
  {}{0pt}{}
\titleformat{\subsection}
  {\normalfont\sffamily\large\bfseries}
  {}{0pt}{}
\titleformat{\subsubsection}
  {\normalfont\sffamily\bfseries}
  {}{0pt}{}
\titleformat*{\paragraph}
  {\sffamily\normalsize}

\usepackage{fancyhdr}
\makeatletter
\let\ps@plain\ps@fancy
\fancyheadoffset[L]{4.5cm}
\fancyfootoffset[L]{4.5cm}

\definecolor{linky}{rgb}{0.0, 0.5, 1.0}

\newtcolorbox{repobox}
   {colback=red, colframe=red!75!black,
     boxrule=0.5pt, arc=2pt, left=6pt, right=6pt, top=3pt, bottom=3pt}

\newcommand{\ExternalLink}{%
   \tikz[x=1.2ex, y=1.2ex, baseline=-0.05ex]{%
       \begin{scope}[x=1ex, y=1ex]
           \clip (-0.1,-0.1)
               --++ (-0, 1.2)
               --++ (0.6, 0)
               --++ (0, -0.6)
               --++ (0.6, 0)
               --++ (0, -1);
           \path[draw,
               line width = 0.5,
               rounded corners=0.5]
               (0,0) rectangle (1,1);
       \end{scope}
       \path[draw, line width = 0.5] (0.5, 0.5)
           -- (1, 1);
       \path[draw, line width = 0.5] (0.6, 1)
           -- (1, 1) -- (1, 0.6);
       }
   }

\patchcmd{\@maketitle}{center}{flushleft}{}{}
\patchcmd{\@maketitle}{center}{flushleft}{}{}
\patchcmd{\@maketitle}{\LARGE}{\LARGE\sffamily}{}{}
\def\maketitle{{%
  
  \AB@maketitle}}
\makeatletter
\renewcommand\AB@affilsepx{ \protect\Affilfont}
\renewcommand\AB@affilnote[1]{{\bfseries #1}\hspace{3pt}}
\renewcommand{\affil}[2][]%
   {\newaffiltrue\let\AB@blk@and\AB@pand
      \if\relax#1\relax\def\AB@note{\AB@thenote}\else\def\AB@note{#1}%
        \setcounter{Maxaffil}{0}\fi
        \begingroup
        \let\href=\href@Orig
        \let\texttt=\textttOrig
        \let\protect\@unexpandable@protect
        \def\thanks{\protect\thanks}\def\footnote{\protect\footnote}%
        \@temptokena=\expandafter{\AB@authors}%
        {\def\\{\protect\\\protect\Affilfont}\xdef\AB@temp{#2}}%
         \xdef\AB@authors{\the\@temptokena\AB@las\AB@au@str
         \protect\\[\affilsep]\protect\Affilfont\AB@temp}%
         \gdef\AB@las{}\gdef\AB@au@str{}%
        {\def\\{, \ignorespaces}\xdef\AB@temp{#2}}%
        \@temptokena=\expandafter{\AB@affillist}%
        \xdef\AB@affillist{\the\@temptokena \AB@affilsep
          \AB@affilnote{\AB@note}\protect\Affilfont\AB@temp}%
      \endgroup
       \let\AB@affilsep\AB@affilsepx
}
\makeatother

\renewcommand\Affilfont{\sffamily\small\mdseries}
\ifnum 0\ifxetex 1\fi\ifluatex 1\fi=0 
  \usepackage[T1]{fontenc}
  \usepackage[utf8]{inputenc}

\else 
  \ifxetex
    \usepackage{mathspec}
    \usepackage{fontspec}

  \else
    \usepackage{fontspec}
  \fi
  \defaultfontfeatures{Ligatures=TeX,Scale=MatchLowercase}

\fi
\IfFileExists{upquote.sty}{\usepackage{upquote}}{}
\IfFileExists{microtype.sty}{%
\usepackage{microtype}
\UseMicrotypeSet[protrusion]{basicmath} 
}{}

\let\addcontentslineOrig=\addcontentsline
\def\addcontentsline#1#2#3{\bgroup
  \let\texttt=\textttOrig\addcontentslineOrig{#1}{#2}{#3}\egroup}
\let\markbothOrig\markboth
\def\markboth#1#2{\bgroup
  \let\texttt=\textttOrig\markbothOrig{#1}{#2}\egroup}
\let\markrightOrig\markright
\def\markright#1{\bgroup
  \let\texttt=\textttOrig\markrightOrig{#1}\egroup}

\usepackage{graphicx,grffile}
\makeatletter
\def\maxwidth{\ifdim\Gin@nat@width>\linewidth\linewidth\else\Gin@nat@width\fi}
\def\maxheight{\ifdim\Gin@nat@height>\textheight\textheight\else\Gin@nat@height\fi}
\makeatother
\setkeys{Gin}{width=\maxwidth,height=\maxheight,keepaspectratio}
\IfFileExists{parskip.sty}{%
\usepackage{parskip}
}{
\setlength{\parindent}{0pt}
\setlength{\parskip}{6pt plus 2pt minus 1pt}
}
\ifx\paragraph\undefined\else
\let\oldparagraph\paragraph
\renewcommand{\paragraph}[1]{\oldparagraph{#1}\mbox{}}
\fi
\ifx\subparagraph\undefined\else
\let\oldsubparagraph\subparagraph
\renewcommand{\subparagraph}[1]{\oldsubparagraph{#1}\mbox{}}
\fi

\DeclareFixedFont{\ttb}{T1}{txtt}{bx}{n}{10} 
\DeclareFixedFont{\ttm}{T1}{txtt}{m}{n}{10}  

\usepackage{color}
\definecolor{deepblue}{rgb}{0,0,0.5}
\definecolor{deepred}{rgb}{0.6,0,0}
\definecolor{deepgreen}{rgb}{0,0.5,0}

\usepackage{listings}

\newcommand\pythonstyle{\lstset{
language=Python,
basicstyle=\scriptsize,
morekeywords={self},              
keywordstyle=\scriptsize\color{deepblue},
emph={MyClass,__init__},          
emphstyle=\ttb\color{deepred},    
stringstyle=\color{deepgreen},
    numbers=left,
frame=tb,                         
showstringspaces=false
}}

\lstnewenvironment{python}[1][]
{
\pythonstyle
\lstset{#1}
}
{}

\input{front_matter/macros}

%% file: front_matter/macros.tex
\newcommand\pythoninline[1]{{\pythonstyle\lstinline!#1!}}

\newcommand{\pkgfont}{\texttt} 
\newcommand{\pkg}[2][]{%
  \if\relax\detokenize{#1}\relax
    \pkgfont{#2}%
  \else
    \href{#1}{\pkgfont{#2}}%
  \fi
}
\newcommand{\derivkit}{\pkg{DerivKit}\xspace}

%% file: front_matter/authorlist.tex
\author[1]{Nikolina \v{S}ar\v{c}evi\'c\thanks{Corresponding author: nikolina.sarcevic@gmail.com}\orcidlink{0000-0001-7301-6415}}
\author[2]{Matthijs van der Wild\orcidlink{0000-0002-3949-3063}}
\author[3]{Cynthia Trendafilova\orcidlink{0000-0001-5500-4058}}

\affil[1]{Department of Physics, Duke University, Science Dr, Durham, NC 27710, USA}
\affil[2]{Department of Physics, Durham University, Lower Mountjoy, South Rd, Durham DH1 3LE, UK}
\affil[3]{Center for AstroPhysical Surveys, National Center for Supercomputing Applications, University of Illinois Urbana-Champaign, Urbana, IL, 61801, USA}

%% file: front_matter/margin.tex
\marginpar{
  \begin{flushleft}
  \sffamily\small

  {\bfseries DOI:} \href{https://doi.org/DOI\_TBD}{\color{linky}{DOI TBD}}

  \vspace{2mm}

  {\bfseries Software}
  \begin{itemize}
    \setlength\itemsep{0em}
    \item \href{N/A}{\color{linky}{Review}} \ExternalLink
    \item \href{https://github.com/derivkit/derivkit}{\color{linky}{Repository}} \ExternalLink
    \item \href{https://derivkit.org}{\color{linky}{Website}} \ExternalLink
  \end{itemize}

  \vspace{2mm}
  \par\noindent\hrulefill\par
  \vspace{2mm}

  {\bfseries Editor:} \href{https://example.com}{Pending editor} \ExternalLink \\
  \vspace{1mm}
  {\bfseries Reviewers:}
  \begin{itemize}
    \setlength\itemsep{0em}
    \item \href{https://github.com/pending}{@pending}
  \end{itemize}

  \vspace{2mm}
  {\bfseries Submitted:} N/A \\
  {\bfseries Published:} N/A

  \vspace{2mm}
  {\bfseries License}\\
  Authors retain copyright and release the work under CC BY 4.0
  (\href{http://creativecommons.org/licenses/by/4.0/}{\color{linky}{link}}).
  \end{flushleft}
}

\vspace{.2cm}

\vspace{.5cm}

%% file: front_matter/abstract.tex
\begin{abstract}
\noindent
\derivkit is a Python package for derivative-based statistical inference.
It implements stable numerical differentiation and derivative assembly utilities for Fisher-matrix forecasting and higher-order likelihood approximations in scientific applications, supporting scalar- and vector-valued models including black-box or tabulated functions where automatic differentiation is impractical or unavailable. 
These derivatives are used to construct Fisher forecasts, Fisher bias estimates, and non-Gaussian likelihood expansions based on the Derivative Approximation for Likelihoods (DALI).
By extending derivative-based inference beyond the Gaussian approximation, \derivkit forms a practical bridge between fast Fisher forecasts and more
computationally intensive sampling-based methods such as Markov chain Monte
Carlo (MCMC).
\end{abstract}

%% file: sections/intro.tex
\section{Statement of Need}
\label{sec:state_of_need}

Reliable numerical derivatives of model predictions with respect to parameters
are a core requirement in many areas of scientific computing, including cosmology, particle physics, and climate science.
In practice, such derivatives are often estimated using fixed-step finite differences even when model evaluations are noisy, irregular, expensive, or available only in tabulated form, leading to fragile or irreproducible inference results.
A representative example can be found in cosmology, where forecasting methods based on the Fisher information matrix \citep{Fisher:1920,Fisher:1922saa,Fisher:1925,Fisher:1935} are popular due to their computational efficiency.
Fisher-based methods rely on first-order derivatives of model predictions with respect to parameters.

These frameworks assume Gaussian posteriors and local linear parameter dependence that can break down in nonlinear regimes \citep{Tegmark_1997,heavens2010statisticaltechniquescosmology}.
Sampling-based methods such as Markov chain Monte Carlo (MCMC), grid sampling, and nested sampling provide robust posterior estimates but scale poorly with dimensionality and model complexity \citep{Christensen:2001gj,Lewis:2002ah,Audren:2012wb,Lewis:2013hha, Tegmark:2000db, Skilling:2004,Skilling:2006gxv,Feroz:2007kg,Feroz:2008xx, Trotta_2008, Verde_2010}.
The Derivative Approximation for Likelihoods (DALI)
\citep{Sellentin_2014,Sellentin:2015axa} extends Fisher formalism by incorporating higher-order derivatives to capture leading non-Gaussian features of the posterior.
Despite its conceptual appeal, DALI has seen limited practical adoption, in part due to the lack of general-purpose software and the numerical challenges associated with computing stable higher-order derivatives.

Automatic differentiation (\pkg{autodiff}) provides exact derivatives for differentiable programs, but is not directly applicable to tabulated models, legacy simulation suites, or workflows involving discontinuities or implicit solvers.
As a result, finite-difference methods remain widely used in scientific software, despite their sensitivity to numerical noise and algorithmic tuning.

\derivkit addresses these challenges by providing a diagnostics-driven framework for computing stable numerical derivatives from general numerical model evaluations, without requiring model rewrites or specialized \pkg{autodiff} frameworks.
While cosmological forecasting serves as a primary motivating example, \derivkit is broadly applicable to inference problems requiring robust derivative-based sensitivity analyses across scientific domains.

%% file: sections/core_functionality.tex
\section{Core Functionality}
\label{sec:core_functionality}

\derivkit is organized into modular components (``kits'') that support derivative-based inference workflows, from numerical differentiation to forecasting and likelihood approximations.

\subsection{DerivativeKit: Numerical derivative engines}

At the core of the library lies \emph{DerivativeKit}, which provides several numerical differentiation strategies that can be selected or combined depending on the numerical properties of the function under consideration.

\paragraph{\textbf{Finite-difference derivatives.}}
\derivkit implements high-order central finite-difference derivatives using 3-, 5-, 7-, and 9-point stencils, supporting derivative orders one through four.
Accuracy and robustness are improved through extrapolation and stabilization techniques such as Richardson extrapolation \citep{10.1098/rsta.1911.0009,10.1098/rsta.1927.0008}, Ridders’ method \cite{Ridders}, and noise-robust Gauss--Richardson schemes \cite{oates2024probabilisticrichardsonextrapolation}.
These methods are computationally efficient and well suited to smooth models with low to moderate numerical noise.

\paragraph{\textbf{Polynomial fitting.}}
For noisy or numerically stiff models, \derivkit provides local polynomial-fit
derivatives \citep{Camera_2016, Euclid2020, Bonici_2023, niko_jmas, Fornberg_2025}.
Two variants are available: a fixed-window polynomial fit, and an adaptive
polynomial-fit method.
The adaptive method constructs domain-aware Chebyshev sampling grids with
automatically chosen scales, applies internal offset scaling and optional ridge
regularization, and may adjust the polynomial degree to improve conditioning.
It reports diagnostics and warns when internal fit-quality checks indicate
unreliable derivative estimates.
Figure~1 shows that, in the presence of noise ($\sigma = 0.2$), adaptive-fit
differentiation yields accurate and precise derivative estimates, whereas
estimator variance limits the reliability of standard finite-difference schemes.

\input{figures/adaptive_demo_ad_fd_compare}

\subsection{CalculusKit: Calculus utilities}

\derivkit provides \emph{CalculusKit}, which builds on the numerical derivative engines.
It constructs common quantities composed of derivatives such as gradients, Jacobians, Hessians, and higher-order derivative tensors for scalar- and vector-valued models.
These utilities provide a consistent interface for assembling derivative structures while delegating numerical differentiation to the underlying derivative engines.

\subsection{ForecastKit: Forecasting and likelihood expansions}

The \emph{ForecastKit} uses derivative and calculus utilities to construct common forecasting quantities including Fisher matrices, Fisher bias estimates, and non-Gaussian likelihood expansions.
In particular, \derivkit provides utilities to assemble the derivative tensors required for the Derivative Approximation for Likelihoods (DALI), enabling practical extensions beyond the Gaussian Fisher-matrix forecasts and direct connections to downstream tasks such as likelihood evaluation, sampling, and visualization.
Some of these functionalities are illustrated in Figure \ref{fig:forecastkit_demo}.

\subsection{LikelihoodKit: Likelihood models}

\derivkit includes a lightweight \emph{LikelihoodKit} that provides basic Gaussian and Poisson likelihood models.
While not intended as a full probabilistic programming framework, these implementations support testing, validation, and end-to-end examples within derivative-based inference workflows.

\subsection{Diagnostics and testing}

\derivkit facilitates user diagnostics by reporting metadata describing sampling geometry, fit quality, and internal consistency checks.
It emits explicit warnings or fallback strategies when tolerance criteria are
not met.

All components are accompanied by extensive unit tests to ensure consistency across derivative methods, calculus utilities, and inference workflows, which is particularly important for numerical differentiation and higher-order derivative handling.

\input{figures/forecast_kit_demo}

%% file: figures/adaptive_demo_ad_fd_compare.tex
\begin{figure}
  \centering
  \begin{minipage}[t]{0.48\textwidth}
    \centering
    \includegraphics[width=\textwidth]{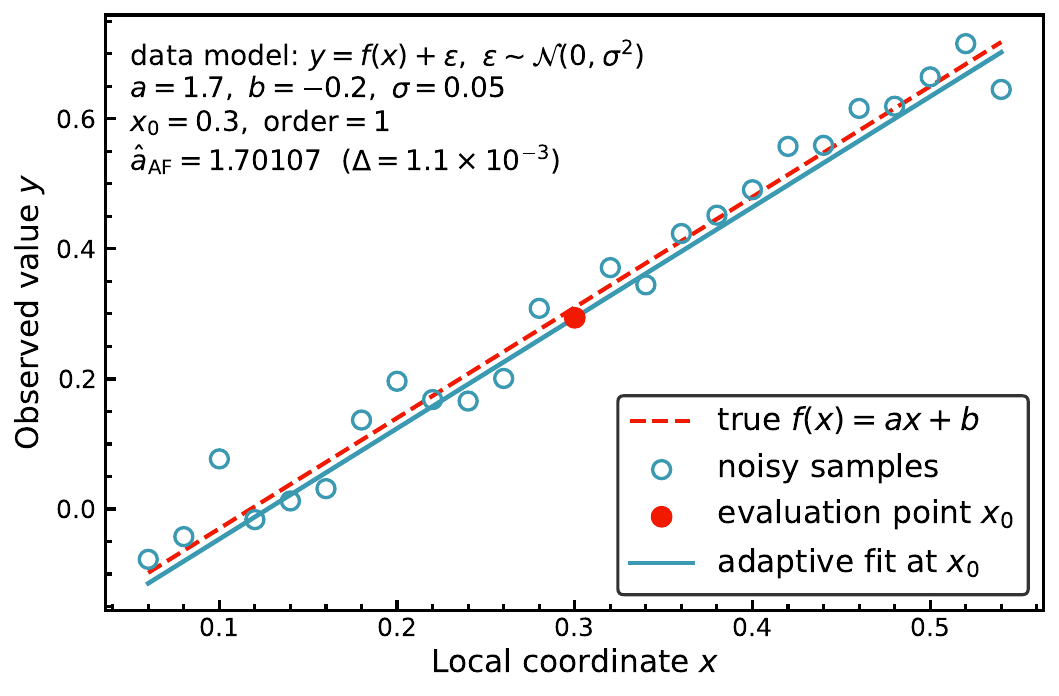}
    \caption*{(a) Adaptive-fit derivative estimation from noisy samples $y=f(x)+\varepsilon$ evaluated near $x_0$.
    A local polynomial fit is used to estimate the derivative at $x_0$
    (blue line); the dashed red curve shows the noise-free function $f(x)$.}
  \end{minipage}\hfill
  \begin{minipage}[t]{0.48\textwidth}
    \centering
    \includegraphics[width=\textwidth]{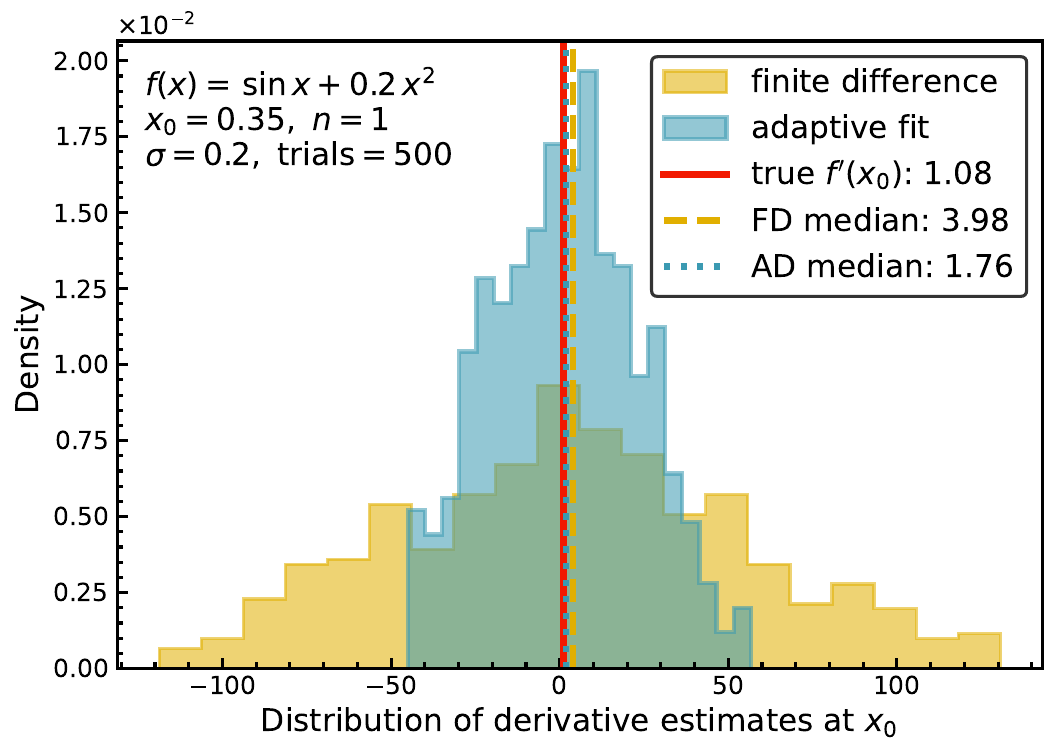}
    \caption*{(b) Distribution of first-derivative estimates at fixed $x_0$ from repeated noise realizations.
    Adaptive-fit and finite-difference stencil methods are compared; vertical lines indicate medians.}
  \end{minipage}

  \caption{
    Adaptive-fit differentiation in the presence of noise.
    Panel (a) illustrates the local fitting procedure.
    Panel (b) shows the improved robustness of the adaptive-fit method relative to standard finite-difference schemes.
  }
  \label{fig:adaptive_demo}
\end{figure}

%% file: figures/forecast_kit_demo.tex
\begin{figure}
  \centering

  \begin{minipage}[t]{0.48\textwidth}
    \centering
    \includegraphics[width=\textwidth]{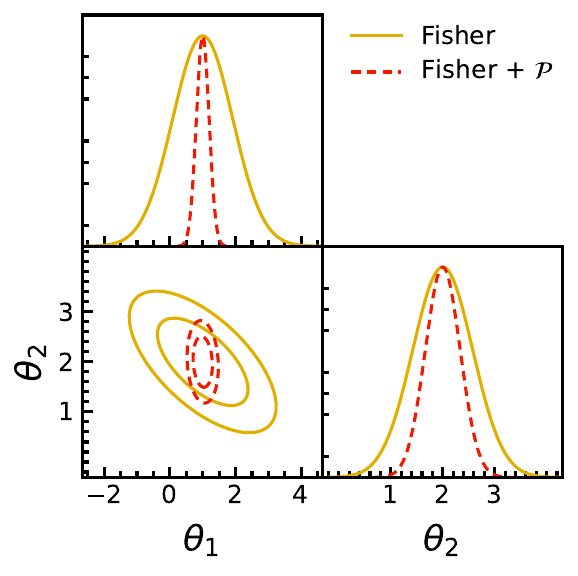}
    \caption*{(a) Standard Fisher contours with and without a Gaussian prior.}
  \end{minipage}\hfill
  \begin{minipage}[t]{0.48\textwidth}
    \centering
    \includegraphics[width=\textwidth]{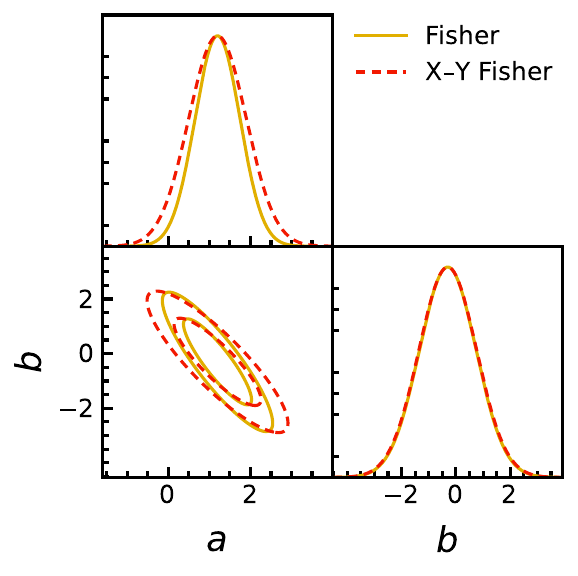}
    \caption*{(b) X--Y Fisher contours accounting for uncertainty in both inputs $x$ and outputs $y$.}
  \end{minipage}

  \vspace{0.6em}

  \begin{minipage}[t]{0.48\textwidth}
    \centering
    \includegraphics[width=\textwidth]{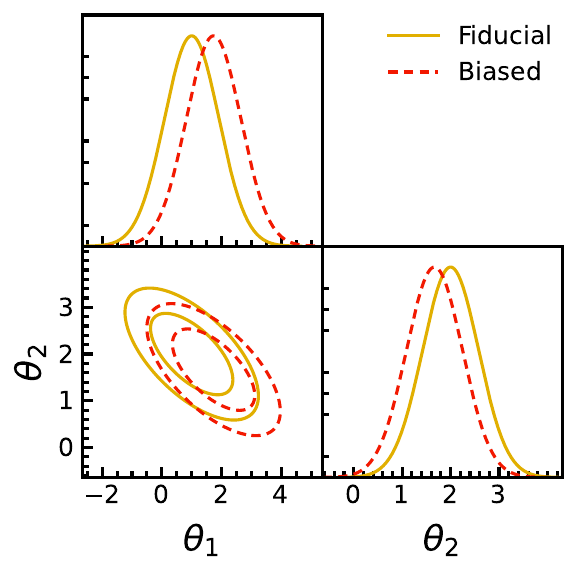}
    \caption*{(c) Fisher bias, showing the parameter shift induced by a biased data vector.}
  \end{minipage}\hfill
  \begin{minipage}[t]{0.48\textwidth}
    \centering
    \includegraphics[width=\textwidth]{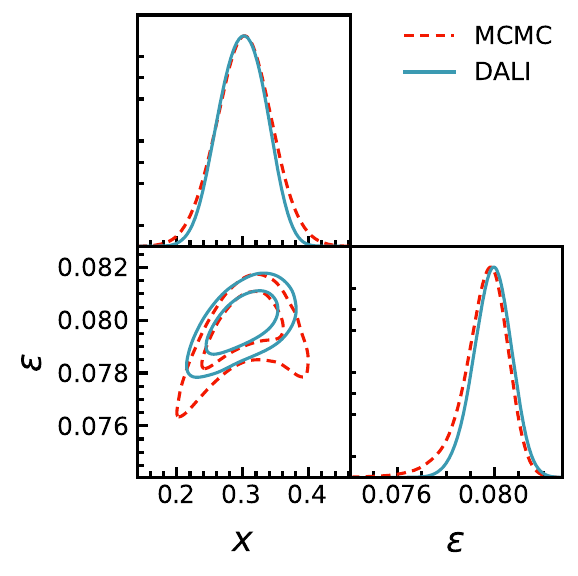}
    \caption*{(d) DALI triplet contours compared to the posterior from MCMC (\pkg{emcee}).}
  \end{minipage}

  \caption{
    Examples of \emph{ForecastKit} functionality.
    Panels show standard Fisher forecasts (with and without priors),
    X--Y Fisher forecasts including uncertainty in both inputs and outputs,
    Fisher bias estimates, and DALI-based non-Gaussian likelihood
    approximations.
  }
  \label{fig:forecastkit_demo}
\end{figure}

%% file: sections/demo.tex
\section{Usage Examples}
\label{sec:usage_examples}

This section presents compact, runnable examples illustrating typical \derivkit workflows.
Additional worked examples and notebooks are available at \url{https://docs.derivkit.org/main/examples/}.

\subsection{DerivativeKit: stable numerical differentiation}
\label{subsec:usage_derivativekit}

The example below compares adaptive polynomial-fit differentiation with a finite-difference baseline for a nonlinear scalar function evaluated at a central point.
The adaptive backend is designed to remain stable in regimes where fixed-step finite differences become sensitive to step-size choice.

\begin{lstlisting}[language=Python, caption={Adaptive derivative estimation compared to finite differences.}, label={lst:usage_derivativekit}]
import numpy as np
from derivkit.derivative_kit import DerivativeKit

def func(x: float) -> float:
    return np.exp(-x*x) * np.sin(3.0*x) + 0.1 * x**3

x0 = 0.3
dk = DerivativeKit(function=func, x0=x0)

d1_adaptive = dk.differentiate(method="adaptive", order=1)

d1_finite = dk.differentiate(
    method="finite",
    order=1,
    extrapolation="ridders")

print("df/dx @ x0:", d1_adaptive, "(adaptive)", d1_finite, "(finite)")
\end{lstlisting}

\subsection{CalculusKit: gradients, Hessians, and Jacobians}
\label{subsec:usage_calculusk}

\derivkit provides calculus utilities for assembling common derivative objects from numerical derivatives.
The example below computes the gradient and Hessian of a scalar-valued model and the Jacobian of a vector-valued model at a central parameter point, delegating numerical differentiation to \emph{DerivativeKit}.

\begin{lstlisting}[language=Python, caption={Calculus objects (gradient, Hessian, Jacobian) with CalculusKit.}, label={lst:usage_calculusk}]
import numpy as np
from derivkit.calculus_kit import CalculusKit

def func_scalar(theta: np.ndarray) -> float:
    x1, x2 = float(theta[0]), float(theta[1])
    return np.exp(x1) * np.sin(x2) \
        + 0.5 * x1**2 * x2**3 \
        - np.log(
            1.0 + x1**2 + x2**2
        )

def func_vector(theta: np.ndarray) -> np.ndarray:
    x1, x2 = float(theta[0]), float(theta[1])
    return np.array([
        np.exp(x1) * np.cos(x2) + x1 * x2**2,
        x1**2 * x2 + np.sin(x1 * x2),
        np.log(1.0 + x1**2 * x2**2) + np.cosh(x1) - np.sinh(x2),
    ], dtype=float)

theta0 = np.array([0.7, -1.2])

ck_scalar = CalculusKit(func_scalar, x0=theta0)
grad = ck_scalar.gradient()
hess = ck_scalar.hessian()
print("grad:", grad)
print("hess:", hess)

ck_vec = CalculusKit(func_vector, x0=theta0)
jac = ck_vec.jacobian()
print("jac shape:", jac.shape)
\end{lstlisting}

\subsection{Forecasting and likelihood expansions}
\label{subsec:usage_forecasting}

Higher-level inference utilities are provided by \emph{ForecastKit}.
The example below constructs a Fisher matrix for a simple two-parameter model, followed by a Fisher bias estimate and the assembly of second-order DALI tensors

\begin{lstlisting}[language=Python, caption={Fisher forecast, Fisher bias, and DALI tensor construction with ForecastKit.}, label={lst:usage_forecastkit_fisher_bias_dali}]
import numpy as np
from derivkit import ForecastKit

def model(theta) -> np.ndarray:
    t1, t2 = float(theta[0]), float(theta[1])
    return np.array([t1 + t2, t1**2 + 2.0 * t2**2], dtype=float)

theta0 = np.array([1.0, 2.0], dtype=float)
cov = np.eye(2, dtype=float)

fk = ForecastKit(function=model, theta0=theta0, cov=cov)

fisher = fk.fisher(method="adaptive")
print("Fisher matrix F:\n", fisher)

# This determines the difference vector
# between the biased and unbiased models
d_unbiased = model(theta0)
d_biased = d_unbiased + np.array([0.5, -0.2], dtype=float) 
delta_nu = fk.delta_nu(data_biased=d_biased, data_unbiased=d_unbiased)

bias_vec, dtheta = fk.fisher_bias(
    fisher_matrix=fisher,
    delta_nu=delta_nu
)
print("Fisher bias vector:\n", bias_vec)
print("Parameter shift:\n", dtheta)

# Construct the second-order DALI corrections
dali_dict = fk.dali()
d1 = dali_dict[2][0]
d2 = dali_dict[2][1]
print("DALI tensors: D1 shape =", d1.shape, ", D2 shape =", d2.shape)
\end{lstlisting}

\derivkit supports Fisher forecasting, Fisher bias calculations, and the construction of higher-order derivative tensors required for DALI likelihood expansions.
These quantities can be used directly for downstream analysis and visualization,
including Gaussian and non-Gaussian posterior contours.
Further scripts and notebook-based demonstrations are available at \url{https://github.com/derivkit/derivkit-demos}.

%% file: sections/use_cases.tex
\section{Use cases}
\label{sec:usecases}

Typical applications of \derivkit span a wide range of scientific and engineering workflows that require robust numerical derivatives of noisy, interpolated, or tabulated model outputs, including surrogate models and emulators.
While cosmological analyses provide a motivating example, the use cases listed below are broadly applicable to any setting where analytic derivatives or reliable automatic differentiation are unavailable.

\begin{enumerate}
  \item \textbf{Fisher forecasts and local sensitivity analyses:}
    Numerical derivatives of model outputs with respect to parameters enter directly into Fisher-matrix forecasts and related local sensitivity measures.
    \derivkit supports diagnostics-driven derivative estimation for noisy, irregular, or computationally expensive models.

  \item \textbf{Higher-order likelihood expansions and non-Gaussian corrections:}
  When posterior distributions deviate from Gaussianity, higher-order derivatives of the log-likelihood are required to capture leading non-Gaussian features.
  \derivkit supports derivative estimation for likelihood expansions such as DALI and related non-Gaussian approximations.

  \item \textbf{Derivative estimation from tabulated or precomputed models:}
  In many simulation pipelines, model predictions are available only on discrete parameter grids or as precomputed lookup tables.
  \pkg{DerivKit} enables stable derivative estimation directly from such tabulated data, avoiding the need for model refactoring or surrogate retraining.

  \item \textbf{Sensitivity studies and gradient validation for black-box models:}
    In workflows involving interpolated models, emulators, or legacy simulation software, analytic gradients are often unavailable or unreliable.
    \derivkit can be used to explore local parameter sensitivities, validate finite-difference or automatic-differentiation gradients, and assess numerical stability.

  \item \textbf{Derivatives of parameter-dependent covariance models:}
    In many inference problems, covariance matrices depend explicitly on model parameters.
    \derivkit supports numerical differentiation of such covariance models, enabling Fisher forecasts and likelihood expansions that consistently account for parameter-dependent noise.

\end{enumerate}

\vspace{0.5em}
A companion study applying \derivkit to standard cosmological probes, including weak lensing and galaxy clustering analyses, cosmic microwave background (CMB) observations, and potentially supernova data, is in preparation (\textit{Šarčević et al.}, in prep.).

%% file: sections/avail.tex
\section{Availability}
\noindent
\textbf{Source:} \href{https://github.com/derivkit/derivkit}{github.com/derivkit/derivkit}. \\
\textbf{License:} MIT (or similar OSI-approved). \\
\textbf{Install:} \texttt{pip\ install\ derivkit} (or \texttt{pip\ install\ -e\ .} for development). \\
\textbf{Examples:}
\href{https://github.com/derivkit/derivkit-demos/}{github.com/derivkit/derivkit-demos}. 

%% file: sections/ack.tex
\section*{Acknowledgments}

N\v{S} is supported in part by the OpenUniverse effort, which is funded by NASA under JPL Contract Task 70-711320, ‘Maximizing Science Exploitation of Simulated Cosmological Survey Data Across Surveys.’
MvdW is supported by the Science and Technology Facilities Council via LOFAR-U.K.~[ST/V002406/1] and UKSRC [ST/T000244/1].
CT is supported by the Center for AstroPhysical Surveys (CAPS) at the
National Center for Supercomputing Applications (NCSA), University of Illinois Urbana-Champaign.
This work made use of the Illinois Campus Cluster, a computing resource that is operated by the Illinois Campus Cluster Program (ICCP) in conjunction with the National Center for Supercomputing Applications (NCSA) and which is supported by funds from the University of Illinois at Urbana-Champaign.
The authors thank Marco Bonici, Bastien Carreres, Matthew Feickert, Arun Kannawadi, Konstantin Malanchev, Vivian Miranda, Charlie Mpetha, and Knut Morå for useful discussions.

\section*{Software Acknowledgments}

This work made extensive use of open-source scientific software.
We acknowledge the Python programming language\footnote{\url{https://www.python.org}}
as the primary development environment.
Core numerical functionality was provided by \pkg{NumPy} \citep{numpy} and
\pkg{SciPy} \citep{scipy}, with visualization handled using
\pkg{Matplotlib} \citep{matplotlib}.
Parallel evaluations within the library are implemented using the
\pkg{multiprocess} package \citep{McKerns2012}.
Interactive development and documentation were supported by
\pkg{Jupyter Notebooks} \citep{jupyter}.
Posterior sampling, analysis, and visualization are handled within the library
via \pkg{emcee} \citep{Foreman_Mackey_2013_emcee} and \pkg{GetDist}
\citep{Lewis_getdist}, which are core dependencies used for sampling,
post-processing, and inference workflows.

\section*{Author Contributions}

N\v{S} initiated and led the project, defined the overall scientific and software
architecture, and was responsible for the majority of the code development,
testing infrastructure, and manuscript preparation.

MvdW made substantial technical contributions throughout the project, including
core code development, extensive code review and refinement, establishment and
enforcement of development standards, expansion of the testing suite, and major
contributions to the user-facing documentation and manuscript.

CT contributed targeted code implementations, performed benchmarking and
validation checks, and provided detailed review and feedback on the software, documentation, and
manuscript.